\documentclass[twocolumn,showpacs,preprintnumbers,amsmath,amssymb,floatfix]{revtex4}

\usepackage{graphicx}
\usepackage{dcolumn}
\usepackage{bm}

\newcommand{\beq}{\begin{equation}}
\newcommand{\eeq}{\end{equation}}
\newcommand{\bey}{\begin{eqnarray}}
\newcommand{\eey}{\end{eqnarray}}

\begin{document}


\title{\underline{Stability of charged thin-shell  wormholes in $(2+1)$ dimensions}}

\author{Ayan Banerjee}
 \email{ayan\_7575@yahoo.co.in}
\affiliation {Department of Mathematics, Jadavpur University,
 Kolkata-700032, India}

\date{\today}

\begin{abstract}
In this paper we construct charged thin-shell wormholes in (2+1)-dimensions
applying the cut-and-paste technique implemented by Visser, from a
 BTZ  black hole which was discovered by Ba\~{n}ados, Teitelboim and Zanelli \cite{BTZ1992},
and the surface stress are determined using the Darmois-Israel formalism
at the wormhole throat.
We analyzed the stability of the shell considering phantom-energy or generalised Chaplygin
gas equation of state for the exotic matter at the throat.
We also discussed the linearized stability of charged
thin-shell wormholes around the static solution.

\end{abstract}


\maketitle

\section{Introduction}
 Last two decades traversable wormholes and thin-shell wormholes are very interesting topic to the scientist,
though it is a hypothetical concept in space-time. At first Morris and Thorns \cite{MT1988} proposed the
structure of traversable wormhole which is the solution of Einstein's field equations having two asymptotically flat  regions connected by
a minimal surface area, called throat, satisfying  flare-out condition \cite{HV1997}. However,
 one has to tolerate the violation of energy condition which is treated as exotic matter.
Visser, Kar and Dadhich \cite{VSD2003} have shown how one can construct a traversable wormhole with small violation of
energy condition.
Different types of  wormholes have been discussed in \cite{
{LFO2003},{L87},{MV},{DSM}}.

~  Though, it is very difficult to deal with the exotic matter
(violation of energy condition), Visser \cite{M.v1989},
 adapted a way to minimize the usage of exotic matter
applying the cut-and-paste technique on a black hole to
construct a new class  of spherically symmetric wormhole,
known as thin-shell wormhole where the  exotic matter
is concentrated at the throat of the wormhole.
The surface stress-energy tensor components, at the throat, are determined
using the Darmois-Israel formalism \cite {IW,PH}.

  It is needed to define an equation of state for the
  exotic matter, located at the throat, though several models
 have been proposed, one of them is the "Phantom-like" equation of state
 defined as: $p$=$w\rho$, $w<-1$, for more specifically
 when $w<-1/3$, then the equation of state is dark energy type
 and when $w>-1/3$, then equation of state is quintessence type.
 Another important equation of state is the "generalised Chaplygin gas",
 defined as: $p$=$-A/\rho^\alpha$, where A is a positive constant
 and $0<\alpha\leq1$.
 A well study for both equation of state on wormholes have been
 discussed in \cite {FSAK,{MPF},{CE},{E},{FSN},{FSN2006},{MOA},{S},{TME},{P}}.

For the local stability of the shell around the static
solution under small perturbation,
Visser and Poission \cite{EP} proposed the linearized stability analysis for thin-shell wormhole  by
joining two Sehwarzschild geometries. Lobo and
Crawford \cite{LC}, extended the idea of linearized
stability analysis with the cosmological constant.Stablity analysis of charged
thin-shell wormholes have been studied in \cite{ECS,EC,{AZF},{FKS}}.
Recently several works have been done in (2+1) dimensions in
\cite{{KJM2004},{Kim2004},{EA},{RS2011},{FR2011},{FAI},{MR},{MJ}}.

 Charged black holes and wormholes are very interesting topic in recent
days. "Charge without charge" effects of Misner-Wheeler \cite{{MC}}  is one
of the most interesting fact produced by wormholes. Morris and Thorns
wormhole takes a new level after the addition of an electric charged
proposed by Kim and Lee \cite{{SS},{SH}}.

In the year 1992, Ba\~{n}ados, Teitelboim and Zanelli (BTZ) \cite{BTZ1992}
 discovered a black hole in (2+1) dimensions with a negative cosmological constant,
which is very similar to (3+1) dimensional black hole. Babichev, Dokuchaev and Eroshenko \cite{EV}
have shown that with accretion of phantom-energy, black hole mass decreases and for the decreasing
mass, the existence of horizon is not crucial. Applying this concept on BTZ black hole,
Jamil and Akbar \cite{MM} have shown that mass evolution is dependent on the pressure and
density of the phantom energy rather than the mass of the black hole.

In this work,  we present a  thin-shell wormhole
from charged  BTZ black holes by
the cut-and-paste technique in (2+1)-dimensions. Here we survey,
 whether the charged thin-shell is stable or not
 when we consider the equation of state is "Phantom-like" or
 "generalised Chaplygin" gas. Linearized stability has
 also discussed around the static solution.
Various aspects of the thin-shell wormholes
are also analyzed.

\section{ construction of  Charged Thin-shell wormhole}
\noindent

 The charged BTZ black hole with a negative cosmological constant
$\Lambda=-1/L{^2}$ is a  solution of (2+1)-dimensional gravity.
The metric is given by \cite{ccj}
\begin{equation}
ds ^2 = - f(r) dt^2 + \frac{dr^2}{f(r)} + r^2 d\phi^2 ,\label{eq22}
\end{equation}
\begin{equation}
f(r)=\frac{r^2}{L{^2}}-\left[M+\frac{Q^2}{2} \ln(\frac{r}{L})\right].
\end{equation}
Where $f(r)$ is known as lapse function. $M$ and $Q$ are mass and
electric charge of the BTZ black hole.
Here we take two identical copies from BTZ black hole with
$r\geq a$:

\[ \mathcal{M}^\pm = ( x \mid r \geq a )  \]

and stick them together at the  junction surface

\[ \Sigma = \Sigma^\pm = ( x \mid r = a )  \]
to get a new geodesically complete manifold
$ \mathcal{M} = \mathcal{M}^+
+\mathcal{M}^- $. The minimal surface area at
$\Sigma$, referred as a throat of wormhole where we
take $ a > r_h $ to avoid the event horizon.

 The junction surface (where the wormhole throat
is located) is one dimensional ring of matter where the
stress energy components are non zero can be evaluated using
the second fundamental forms\cite{{MR},{MJ},{FAI},{EA}} defined by

\begin{equation}K^{i\pm}_j =  \frac{1}{2} g^{ik}
\frac{\partial g_{kj}}{\partial \eta}   \Bigl\lvert_{\eta =\pm 0} =
\frac{1}{2}   \frac{\partial r}{\partial \eta} \Bigl\lvert_{r=a}
 ~ g^{ik} \frac{\partial g_{kj}}{\partial r}\Bigl\rvert_{r=a}.
\label{eq36}
\end{equation}
 Here, $\eta$ is the Riemann normal coordinates, positive and
negative in two side of the boundaries with $x^\mu = ( \tau,\phi,\eta) $,
where $\tau$ represents the proper time on the shell.

Since, $K_{ij}$ is not continuous for the shell around $\Sigma$, so the second fundamental forms with
 discontinuity are given by
\begin{equation}\mathcal{K} _{ij} =   K^+_{ij}-K^-_{ij}.
\label{eq37}
 \end{equation}

Now, one can define the surface stress energy tensor $S^i_j$, at the
junction. Using Lanczos equation follow from the Einstein equations lead to
 \begin{equation}S^i_j=-\frac{1}{8\pi}\left(K^i_j-\delta^i_jK^k_k\right).
\label{eq37}
 \end{equation}

 Considering circular symmetry,  $K^i_j$ becomes
\begin{equation}
K^i_j =
\left( {\begin{array}{cc}
 K^\tau_\tau & 0  \\
 0 & K^\phi_\phi \\
 \end{array} } \right),\label{37}
\end{equation}
then the surface stress-energy tensor become
\begin{equation}
S^i_j =
\left( {\begin{array}{cc}
 -\sigma & 0  \\
 0 & -v \\
 \end{array} } \right),\label{38}
\end{equation}
   where $\sigma$ is surface energy density
and $v$ is surface pressure, then Lanczos equation becomes

\begin{eqnarray} \sigma &=&  -\frac{1}{8\pi}  \kappa _\phi^\phi,\\
\label{eq38} v &=&  -\frac{1}{8\pi}  \kappa _\tau^\tau.
\label{eq39}
 \end{eqnarray}

Now, using the Eqs.(3-9) for the metric given in Eq.(1), we get the above expression
  as
 \begin{equation}\sigma =  -\frac{1}{4\pi a}  \sqrt{ \frac{a^2}{L{^2}}-\left[M+\frac{Q^2}{2} \ln(\frac{a}{L})\right]+\dot{a}^2
 },\end{equation}
 \begin{equation} p = -v =  \frac{1}{4\pi}  \left[ \frac{\frac{a}{r_0{^2}}-\frac{Q^2}{4a} + \ddot{a}}
{\sqrt{\frac{a^2}{L{^2}}-\left[M+\frac{Q^2}{2} \ln(\frac{a}{L})\right]+\dot{a}^2}}  \right].
\end{equation}

 Here, overdot  denotes the derivatives with
respect to $\tau$, assuming  the
throat radius is a function of proper time i.e. $ a = a(\tau)$.

 For the static configuration of radius $a$, we have (i.e.
$\dot{a} = 0 $ and $\ddot{a}= 0 $) from  Eqs.(10) and (11)

\begin{equation}\sigma =  -\frac{1}{4\pi a}  \sqrt{ \frac{a^2}{L{^2}}-\left[M+\frac{Q^2}{2} \ln(\frac{a}{L})\right]
 },
\end{equation}
\begin{equation} p = -v =  \frac{1}{4\pi}  \left[ \frac{\frac{a}{L{^2}}-\frac{Q^2}{4a} }
{\sqrt{\frac{a^2}{L{^2}}-\left[M+\frac{Q^2}{2} \ln(\frac{a}{L})\right]}}  \right].
\end{equation}

 The energy condition demands, if $\sigma \geq 0$ and $ \sigma+ p\geq 0$ are
satisfied, then the weak energy condition (WEC) holds and by continuity, if $ \sigma+ p\geq 0$ is
satisfied, then the null energy condition (NEC) holds. Moreover, the strong energy(SEC) holds,
if $ \sigma+ p\geq 0$ and $ \sigma+ 2p\geq 0$ are satisfied. We get from
 Eq.(12) and (13), that $\sigma< 0$ but $ \sigma+ p\geq 0$ and
 $ \sigma+ 2p\geq 0$, for all values of $q$, $M$ and a, which show that the
shell contains matter, violates the weak energy condition
and obeys the null and strong energy conditions which is shown in Fig.(1).
\begin{figure}[ptb]
\begin{center}
\vspace{0.2cm}
\includegraphics[width=0.55\textwidth]{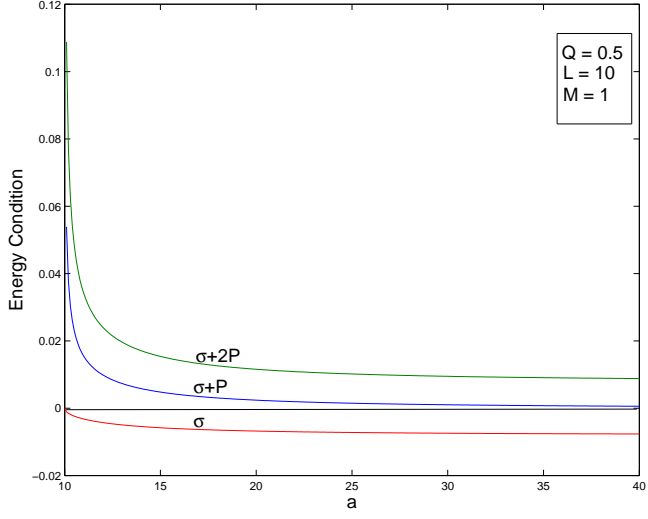}
\end{center}
\caption{Variations of the
expressions of energy conditions
shown against a.  } \label{fig5}
\end{figure}

      Using different values of mass $(M)$ and charge $(Q)$, we
 plot  $\sigma$ and $p$ as a function of the a, shown in Figs.2-5.

\begin{figure}[ptb]
\begin{center}
\vspace{0.3cm}
\includegraphics[width=0.55\textwidth]{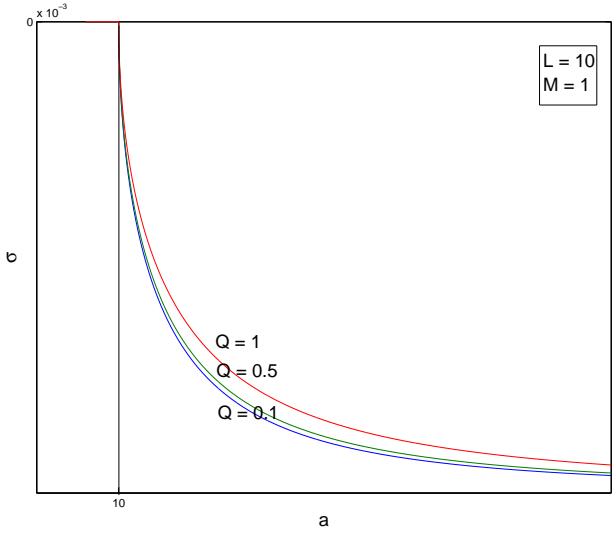}
\end{center}
\caption{Plot for $\sigma$ versus a: Different values of charge(Q) when mass$(M=1)$ is fixed.   } \label{fig5}
\end{figure}

\begin{figure}[ptb]
\begin{center}
\vspace{0.3cm}
\includegraphics[width=0.55\textwidth]{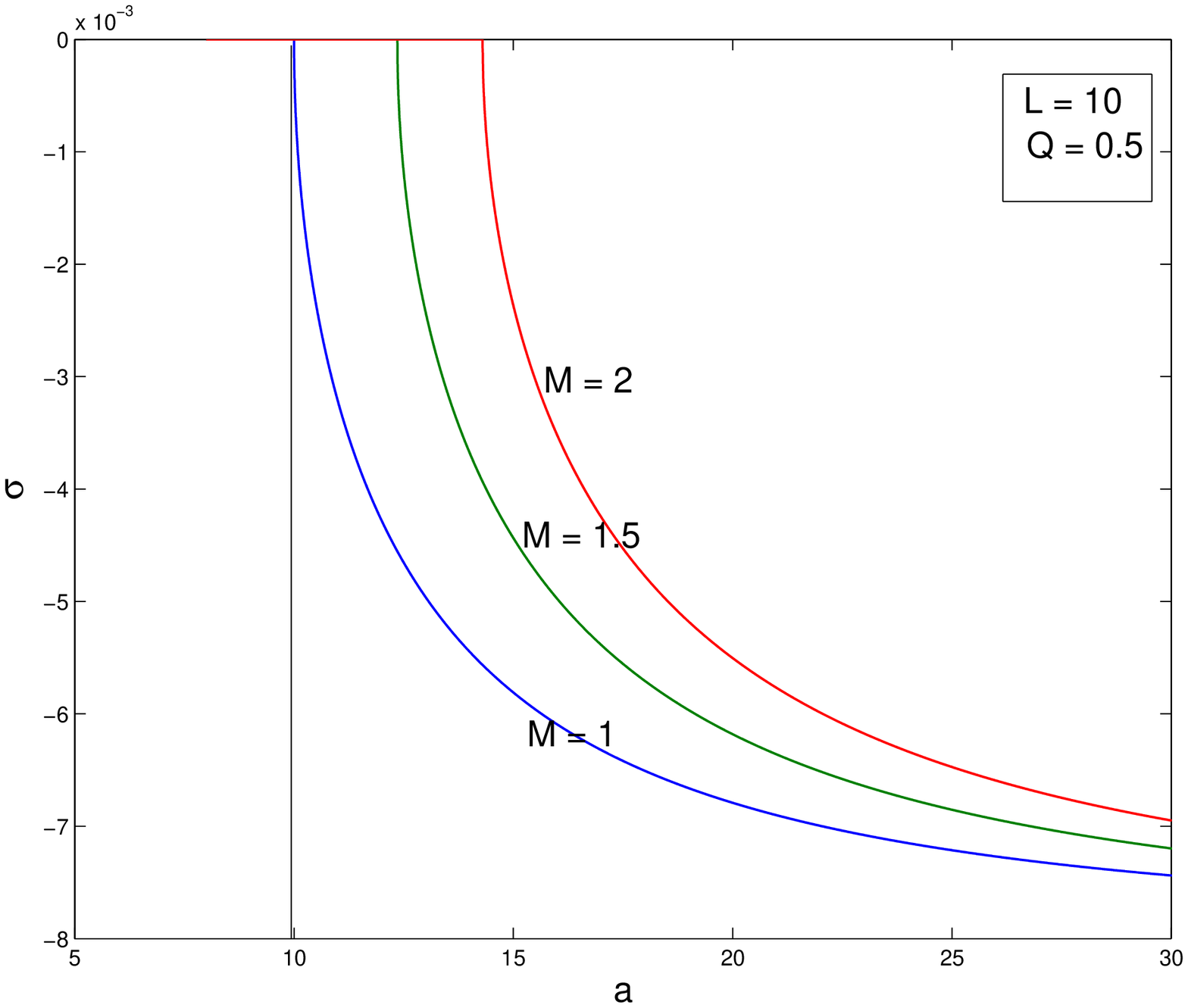}
\end{center}
\caption{Plot for $\sigma$ versus a: Different values of mass$(M)$  when charge(Q=0.5) is fixed.   } \label{fig5}
\end{figure}

\begin{figure}[ptb]
\begin{center}
\vspace{0.3cm}
\includegraphics[width=0.55\textwidth]{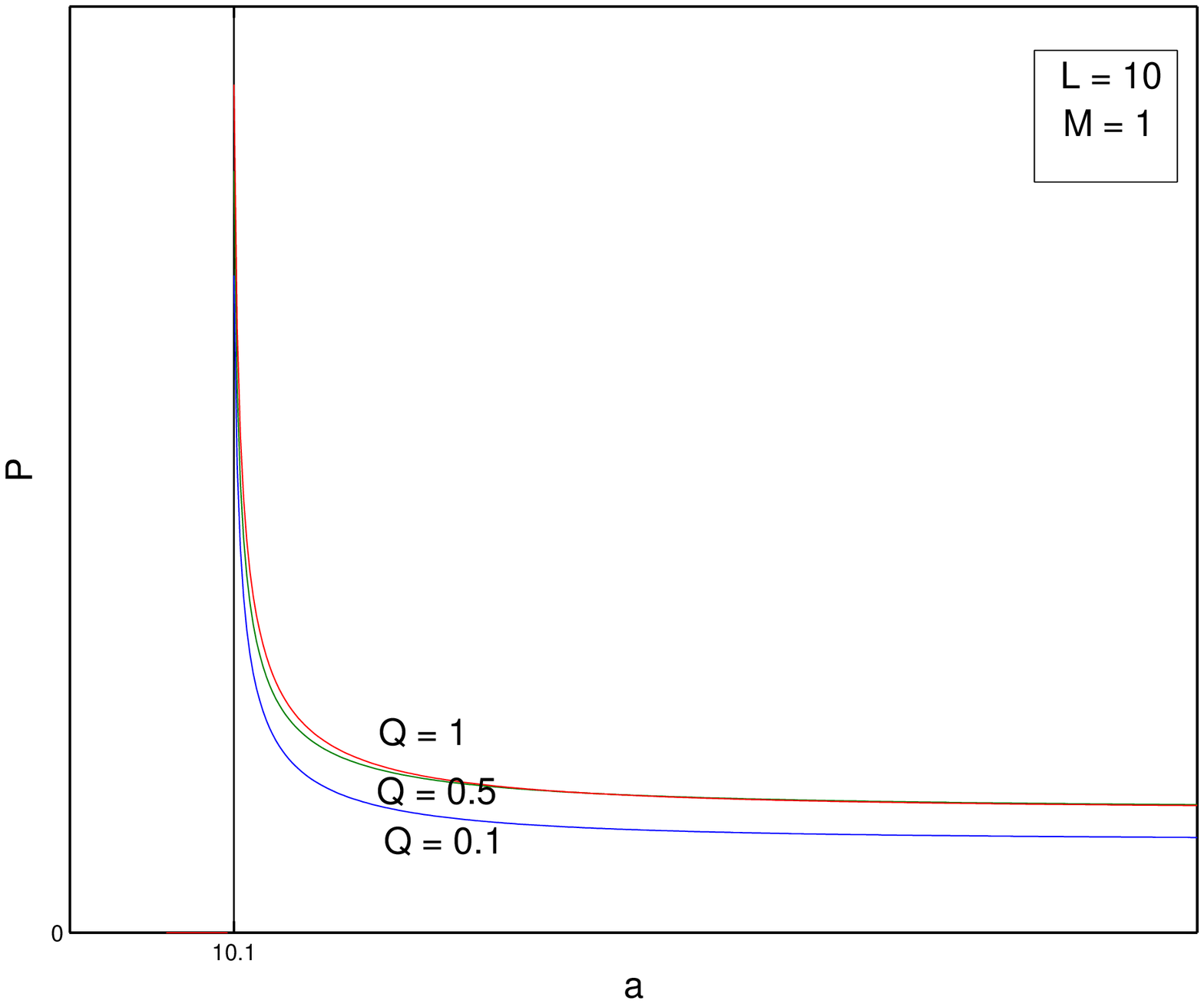}
\end{center}
\caption{Plot for $p$ versus a: Different values of charge(Q) when mass$(M=1)$ is fixed. } \label{fig5}
\end{figure}
\begin{figure}[ptb]
\begin{center}
\vspace{0.3cm}
\includegraphics[width=0.55\textwidth]{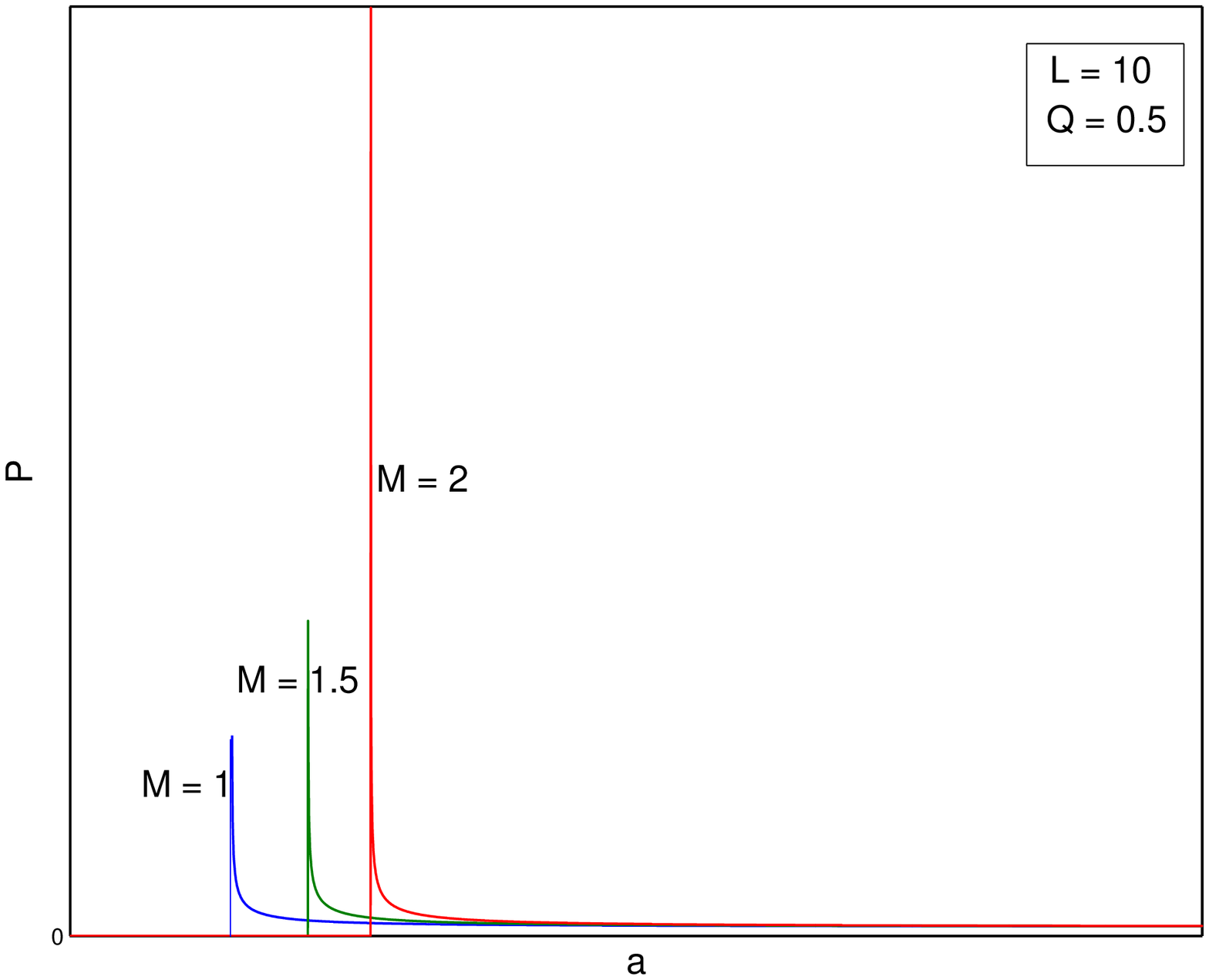}
\end{center}
\caption{Plot for $p$ versus a: Different values of mass$(M)$  when charge(Q=0.5) is fixed.} \label{fig5}
\end{figure}

\section{The gravitational field}

In this section we  analyze the  attractive and repulsive nature of
the wormhole, constructed from charged BTZ black hole and calculate the
observer's three-acceleration $a^\mu = u^\mu_{\,\,;\nu} u^\nu$,
where $u^{\nu} =  d x^{\nu}/d {\tau} =(1/\sqrt{f(r)},
0,0)$. Only non-zero component for the line element in Eq.(1),
is given by

\begin{equation}
a^r = \Gamma^r_{tt} \left(\frac{dt}{d\tau}\right)^2 \\= \frac{r}{L{^2}}-\frac{Q^2}{4r}.
\end{equation}

    A  test particle when radially moving and  initially at rest,
obeys the equation of motion
\begin{equation}
\frac{d^2r}{d\tau^2}= -\Gamma^r_{tt}\left(\frac{dt}{d\tau}
\right)^2 =-a^r,
\end{equation}

which gives the geodesic equation if $a^r=0$. Also, we observe
 that the wormhole is attractive when $a^r> 0$ and repulsive when $a^r < 0$,
 which is shown in Fig.6
\begin{figure}[ptb]
\begin{center}
\vspace{0.6cm}
\includegraphics[width=0.52\textwidth]{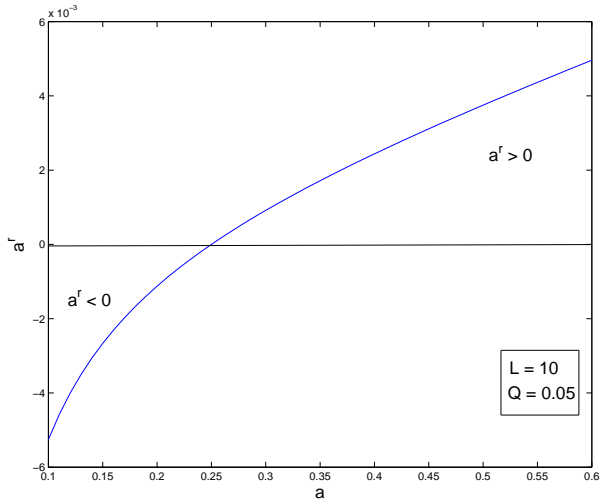}
\end{center}
\caption{ The wormhole is attractive when $a^r>0$ and repulsive when $a^r < 0$.   } \label{fig5}
\end{figure}

\section{The total amount of exotic matter}
To construct such a thin-shell wormhole, we need
exotic matter. Though, using BTZ in the
thin-shell wormhole construction is that, it is not
asymptotically flat and therefore the wormholes are not
asymptotically flat. Recently Mazharimousavi, Halilsoy and Amirabi \cite{{MH}}
shown  that a non-asymptotically
flat black hole solution provides Stable  thin-shell wormholes
which are entirely supported by exotic matter.

In this section, we evaluate the total
 amount of exotic matter for the shell which can
be quantified by the integral \cite{{KY},{ES},{MC},{FK}}

\begin{equation}
   \Omega_{\sigma}=\int [\rho+p_r]\sqrt{-g}d^2x,
\end{equation}
where $g$ represents the determinant of the metric tensor.
Now, by using the radial coordinate $R=r-a$, we have
\begin{equation}
 \Omega_{\sigma}=\int^{2\pi}_0 \int^{\infty}_{-\infty}
     [\rho+p_r]\sqrt{-g}\,dR d\phi.
\end{equation}
For the infinitely thin shell it does not exert any radial
pressure i.e. $p_r=0$ and using  $\rho=\delta(R)\sigma(a)$ for above integral
we have
\begin{multline}\label{E:amount}
 \Omega_{\sigma}=\int^{2\pi}_0 \left.[\rho\sqrt{-g}]
   \right|_{r=a} d\phi=2\pi a\sigma(a)\\
=- \frac{1}{2}\sqrt{ \frac{a^2}{L{^2}}-\left[M+\frac{Q^2}{2} \ln(\frac{a}{L})\right]}.
\end{multline}
With the help of graphical representation (Fig.7), we are trying to describe
the variation of the total amount of
exotic matter on the shell with respect to the  mass and  the charge.

\begin{figure}[ptb]
\begin{center}
\vspace{0.6cm}
\includegraphics[width=0.55\textwidth]{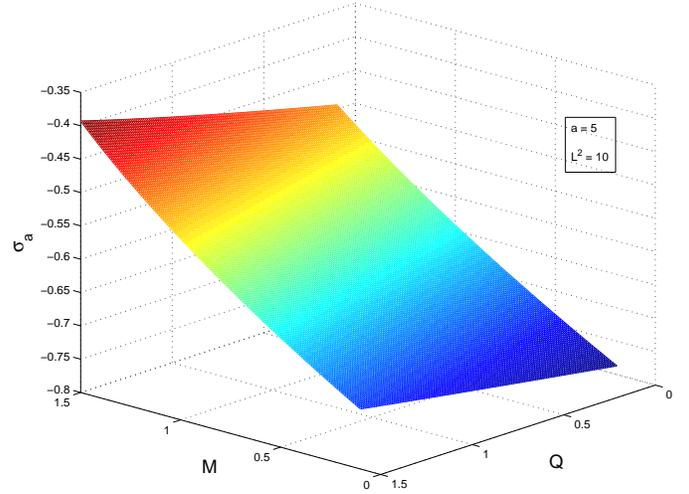}
\end{center}
\caption{variation of the total amount of
exotic matter on the shell with respect to the  mass and  the charge.} \label{fig5}
\end{figure}

\section{Stability}

Stability is one of the important issue for the wormhole.
Here we analyze the stability of the shell from various
angle.Our approaches are (i) phantom-like equation of state
(ii)generalized chaplygin gas equation of state and (iii) linearized radial
perturbation, around the static solution.
\subsection{phantom-like equation of state}
Here, we are trying to describe the stability of the shell
considering the equation of state when the surface energy density and the
surface pressure are taken into account. We set an equation $w=p/\sigma$ i.e. $p=w\sigma$
known as Phantom-like equation of state when  $w<0$.
Since, it is easy to check that the surface pressure and energy density obey the conservation equation
\begin{equation} \frac{d(a \sigma )}{d\tau}+p\frac{d}{d\tau}(a)=0 ,\end{equation}
after differentiating w.r.t $\tau$, one can get
\begin{equation}
\frac{\dot{\sigma}}{\dot{a}}+\frac{1}{a}\left(\sigma+p\right)=0,
\end{equation}
and using $p$=$w\sigma$, we get
\begin{equation}
\frac{d\sigma}{da}+\frac{1}{a}\sigma(1+w)=0.
\end{equation}
Now, consider the static solution with radius $a=a_0$, we have
\begin{equation}  \sigma(a) =\sigma(a_0)\left(\frac{a_0}{a}\right)^{(1+w)} , \end{equation}
 rearranging  the Eq.(10) we can write \begin{equation}  \dot{a}^2+V(a)=0 .  \end{equation}
Which is the equation of motion of the shell, where the potential V(a) is defined as

\begin{equation} V(a)=f(a)-16\pi^2 a^2\sigma^2(a) .\end{equation}
 Substitute the value of $\sigma(a)$ from Eq.(22), one can get

\begin{equation}
V(a)=\frac{a^2}{L{^2}}-\left[M+\frac{Q^2}{2}\ln(\frac{a}{L})\right]-16\pi^2a^2\sigma_0^2\left(\frac{a_0}{a}\right)^{2(1+w)},
\end{equation}where $\sigma_0=\sigma(a_0)$, and rewriting the above equation we have
\begin{equation}
V(a)=\frac{a^2}{L{^2}}-\left[M+\frac{Q^2}{2}\ln(\frac{a}{L})\right]-16\pi^2a^{-2w}\mathcal{A}.
\end{equation}
With $\mathcal{A}=\sigma^2_0a_0^{2(1+w)}$.
Expanding $V(a)$ around the static solution i.e. at $a=a_0$,
we have
\begin{eqnarray}
V(a) &=&  V(a_0) + V^\prime(a_0) ( a - a_0) +
\frac{1}{2} V^{\prime\prime}(a_0) ( a - a_0)^2  \nonumber \\
&\;& + O\left[( a - a_0)^3\right],
\end{eqnarray}
where the primes denote the derivative with respect to a.
The wormhole is stable if and only if $V(a_0)$ has local
minimum at $a_0$ and $V^{\prime\prime}(a_0)>0$. Therefore,
using the conditions $V(a_0)=0$ and $V^{\prime}(a_0)=0$,
with $V^{\prime\prime}(a_0)$ is given by
\begin{equation}
V^{\prime\prime}(a_0)=\frac{2}{L{^2}}+\frac{Q^2}{2a^2_0}-32\pi^2\sigma^2_0w\left(2w+1\right).
\end{equation}
  We have verified that the inequality $V^{\prime\prime}(a_0)>0$,
holds for suitable choice of $Q$, $M $ and $L$ when $w$ takes the
different values. We are trying to describe the stability
of the configuration with help of graphical representation.
In Figs.8-10, we plot the graphs to find the possible range
of $a_0$ where $V(a_0)$ possess a local minimum. Thus, fixing
the values of $Q= 0.5$, $M= 1 $ and $L= 10$, we plot three
different graphs (Figs.8-10), corresponding to the Eq.(28)
 for different values of
$w = -0.1, -0.4, -2.5$ respectively, which represents
different equation of state.
In other words, we get stable thin-shell wormholes from
charged BTZ black hole for different
values of $w$.

\begin{figure}[ptb]
\begin{center}
\vspace{0.6cm}
\includegraphics[width=0.55\textwidth]{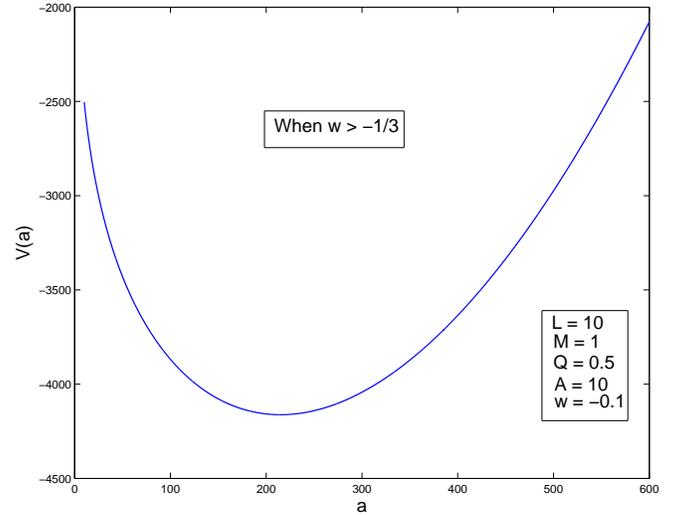}
\end{center}
\caption{Stable wormhole when $w>-1/3$ i.e.
when equation of quintessence state.} \label{fig5}
\end{figure}

\begin{figure}[ptb]
\begin{center}
\vspace{0.6cm}
\includegraphics[width=0.55\textwidth]{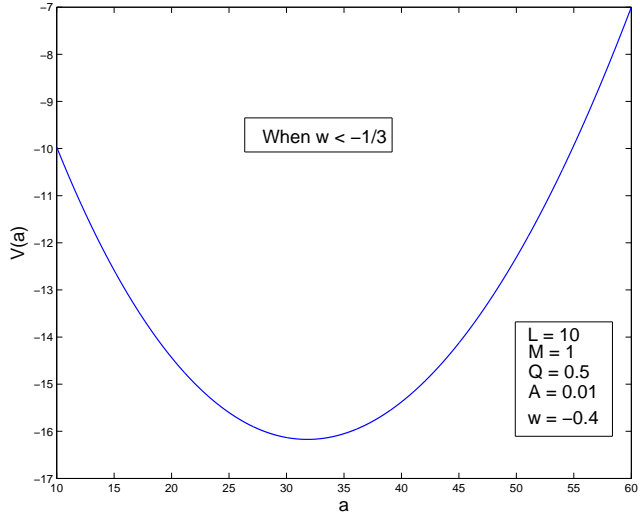}
\end{center}
\caption{Stable wormhole when $w<-1/3$ i.e.
when equation of dark energy state. } \label{fig5}
\end{figure}
\begin{figure}[ptb]
\begin{center}
\vspace{0.6cm}
\includegraphics[width=0.55\textwidth]{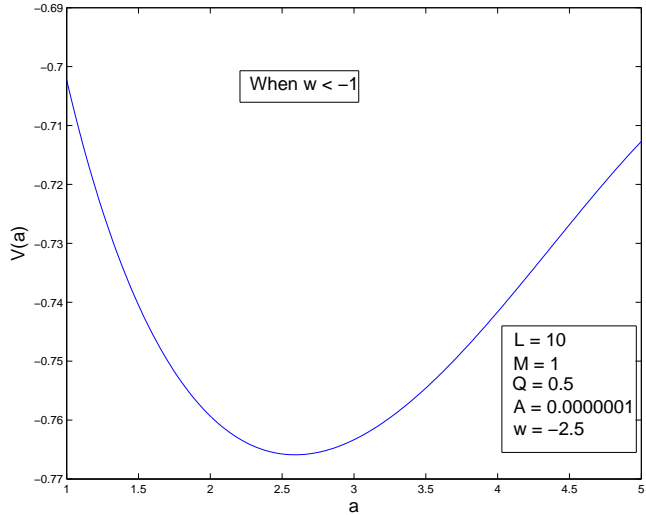}
\end{center}
\caption{Stable wormhole when $ w<-1$ i.e.
when equation of phantom energy state.} \label{fig5}
\end{figure}
\subsection{Generalized Chaplygin gas equation
of state}
   Here we are trying to check the stability
 of the shell considering Chaplygin gas equation of state,
 at the throat, which is a hypothetical substance satisfying an equation of state:
\begin{equation} p=- \frac{A}{\sigma^\alpha} ,  \end{equation}
where $\sigma$ is surface energy density and $p$ is surface pressure
 with  $  A  $ positive constants and $0<\alpha \leq 1$.

substitute the values of $\sigma$ and $p$ (i.e. equations (10) and (11)) in equation (29), we get

\begin{equation}
\frac{(-1)^\alpha\left[\frac{a}{L{^2}}-\frac{Q^2}{4a}+\ddot{a}\right]}{(4\pi)^{\alpha+1}a^\alpha} \left[\frac{a^2}{L{^2}}-[M+\frac{Q^2}{2}\ln(\frac{a}{L})]+\dot{a}^2\right]^\frac{\alpha-1}{2}+A =0   \end{equation}

     For the case of static solution, the surface energy density and surface pressure are given by the equations
(12) and (13), and relate to the equation(29), one can get the solution as

\begin{equation}  \frac{(-1)^\alpha}{(4\pi)^{\alpha+1}a^\alpha}\left[\frac{a}{L{^2}}-\frac{Q^2}{4a}\right] \left[\frac{a^2}{L{^2}}-[M+\frac{Q^2}{2}\ln(\frac{a}{L})]\right]^\frac{\alpha-1}{2}+A =0 . \end{equation}

By assuming L.H.S of Eq.(31) as $\mathcal{G}(a)=0$, we get the  throat radius of the shell at some $a=a_0$.
In Fig(11), we show that $\mathcal{G}(a)$ cuts the a-axis at
$ a_0 = 2.8$ for the value of $\alpha = 0.2$, which represents
 the radius of the throat of shell.

Now, we return to the conservation Eq.(20) and using Eq.(29), we have
\begin{equation}
\dot{\sigma}+\frac{\dot{a}}{a}\left(\frac{\sigma^{\alpha+1}-A}{\sigma^{\alpha}}\right)=0,
\end{equation}
after solving the above equation, one can get
\begin{equation}
              \sigma =\left[A+\left(\sigma_0^{\alpha+1}-A\right)\left(\frac{a_0}{a}\right)^{\alpha+1}\right]^\frac{1}{\alpha+1}.
               \end{equation}

where $\sigma_0$=$\sigma(a_0)$.
Therefore potential $V(a)$, as defined in Eq.(24) takes of the form
\begin{equation}
        V(a)=f(a)-16\pi^2a^2\left[A+\left(\sigma_0^{\alpha+1}-A\right)\left(\frac{a_0}{a}\right)^{\alpha+1}\right]^\frac{2}{\alpha+1}.
               \end{equation}
Where   \begin{equation}
f(a)=\frac{a^2}{L{^2}}-\left[M+\frac{Q^2}{2} \ln(\frac{a}{L})\right].
\end{equation}
 The above solution gives a stable configuration if the
second order derivative of the potential is positive for
the static solution and $V(a_0)$ posses a local minimum at $a_0$.
To find the range of $a_0$ for which $V^{\prime\prime}(a_0)> 0$ we use
graphical representation due to
complexity of the expression $V^{\prime\prime}(a_0)$
for the set of parameters in Eq.(34). Therefore, corresponding
to the radius of the throat at $a_0= 2.8$ (Fig.11), we
are trying to find the possible range of $a_0$ for
which $V(a_0)$ posses a local minima. In Fig.12
we show the range of $a_0$ for the value of $\alpha = 0.2$
and other corresponding parameters (values are given
in the caption of the Fig. 12), for the stable configuration.
Thus, we can get stable thin-shell wormholes
supported by exotic matter filled with Chaplygin gas
equation of state.

\begin{figure}[ptb]
\begin{center}
\vspace{0.6cm}
\includegraphics[width=0.55\textwidth]{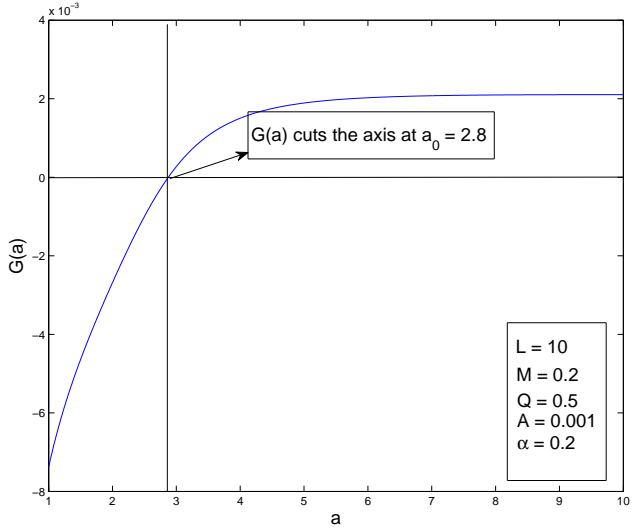}
\end{center}
\caption{ radius of the throat when $\mathcal{G}(a)=0$ cuts the axis a at some $a=a_0$.
} \label{fig5}
\end{figure}

\begin{figure}[ptb]
\begin{center}
\vspace{0.6cm}
\includegraphics[width=0.55\textwidth]{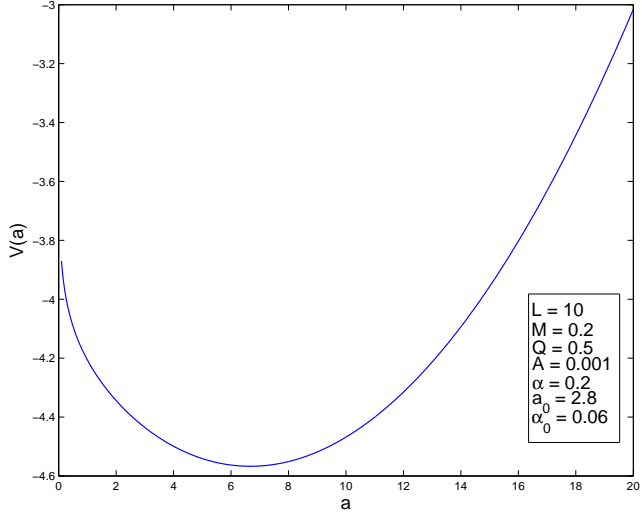}
\end{center}
\caption{The stable wormhole for $0<\alpha\leq1$ when $V(a)$ varies w.r.t a.
} \label{fig5}
\end{figure}
\subsection{Linearized stability
analysis}
Now we are trying to find the stability of the configuration around the
static solution at $a=a_0$ under radial
perturbation. Applying the Taylor series expansion for the potential $V(a)$ around $a_0$,
we get
\begin{eqnarray}
V(a) &=&  V(a_0) + V^\prime(a_0) ( a - a_0) +
\frac{1}{2} V^{\prime\prime}(a_0) ( a - a_0)^2  \nonumber \\
&\;& + O\left[( a - a_0)^3\right],
\end{eqnarray}
where the prime denotes $d/da$.  After differentiating Eq. (24), we obtain

\begin{equation}
V^\prime(a)=\frac{2a}{L{^2}}-\frac{Q^2}{2a}-32\pi^2a\sigma\left(\sigma+a\sigma^\prime\right).
\end{equation}

Now from Eq.(20), we can write $\sigma^\prime$=$-\frac{1}{a}\left(\sigma+p\right)$
where $\sigma^\prime$=$\dot{\sigma}/\dot{a}$ and using Eq(37) we get
\begin{equation}
V^\prime(a)=\frac{2a}{L{^2}}-\frac{Q^2}{2a}+32\pi^2a p\sigma.
\end{equation}
Again differentiating w.r.t a and using the parameter
$ \beta^2(\sigma)$ =$\left. \frac{ \partial
p}{\partial \sigma}\right\vert_\sigma$,
one can get

\begin{equation}
       V^{\prime\prime}(a) = \frac{2}{L{^2}}+\frac{Q^2}{2a^2}-32\pi^2p^2-32\pi^2\beta^2(p\sigma+\sigma^2).
               \end{equation}

Since we are linearizing around $ a = a_0 $, we require that $ V(a_0) =
0 $ and $ V^\prime(a_0)= 0 $. Now using the values of $\sigma$ and $p$ from Eqs.(12) and (13), we get the
 solution of $ V^{\prime\prime}(a_0) $ given by

\begin{equation}
V^{\prime\prime}(a_0)=\frac{2}{L{^2}}+\frac{Q^2}{2a_0^2}-\frac{2\beta^2}{a_0^2}\left[\frac{Q^2}{4}-H\right]-2\frac{\left[\frac{a_0}{L{^2}}-\frac{Q^2}{4a_0}\right]^2}
{\frac{a_0^2}{L{^2}}-H},
\end{equation}
where $ H = M+\frac{Q^2}{2}\ln(\frac{a_0}{L}).$

The configuration is in stable equilibrium if and only if $  V^{\prime\prime}(a_0) >0$. So starting with $ V^{\prime\prime}(a_0)=0$
and solve for $\beta^2 $ we get
\begin{equation}
{\beta_0}^2=\mathcal{X}_0(a_0)\equiv1+\frac{8a_0^2Q^2-\left(L{^2}Q^4+16H^2L{^2}\right)}{4a_0^2Q^2-4H\left(4a_0^2+L{^2}Q^2\right)+16H^2L{^2}}.
 \end{equation}
Here we observe that

\[(i)~~ {\beta_0}^2<\mathcal{X}_0(a_0),  ~~ if ~~   4a_0^2Q^2>L{^2}\left(Q^4+16H^2\right),\]
 \[(ii)~~ {\beta_0}^2>\mathcal{X}_0(a_0),  ~~ if ~~   4a_0^2Q^2<L{^2}\left(Q^4+16H^2\right).\]
 The value of ${\beta_0}^2$  which represent the velocity of sound for ordinary
matter, does not exceed the speed of light and
lies in the region of $0 \leq{\beta_0}^2 < 1$.
In Fig.13, the region of stability depicted below the
solid line, corresponding to the values of $M =1$, $Q = 0.5$, and $L^2 = 0.01$,
where in Figs.(14-15), the region of stability depicted
above the left and below the right of the solid
line, corresponding to the value of $M =1$, $L^2 = 10$ and increasing
the value of charge Q, respectively. In all cases the
stability region of ${\beta_0}^2$ is grater than one.
As we are dealing with exotic matter, we relaxed the range of
${\beta_0}^2$ for ordinary matter in
our stability analysis for the thin-shell wormholes.

\begin{figure}[ptb]
\begin{center}
\vspace{0.6cm}
\includegraphics[width=0.53\textwidth]{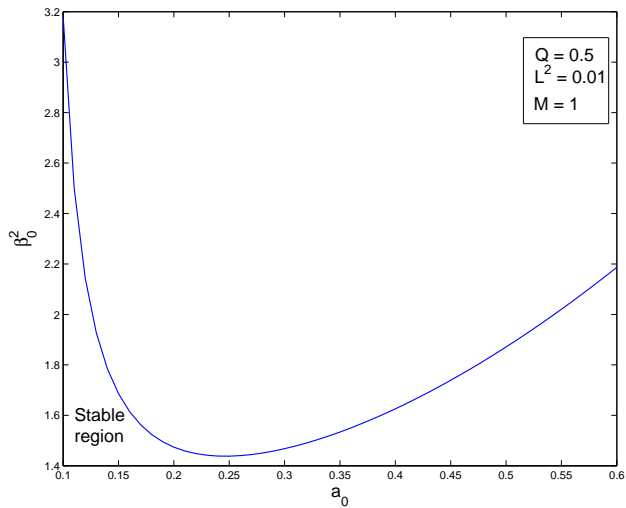}
\end{center}
\caption{ The region of stability is below the curve.
} \label{fig5}
\end{figure}
\begin{figure}[ptb]
\begin{center}
\vspace{0.4cm}
\includegraphics[width=0.55\textwidth]{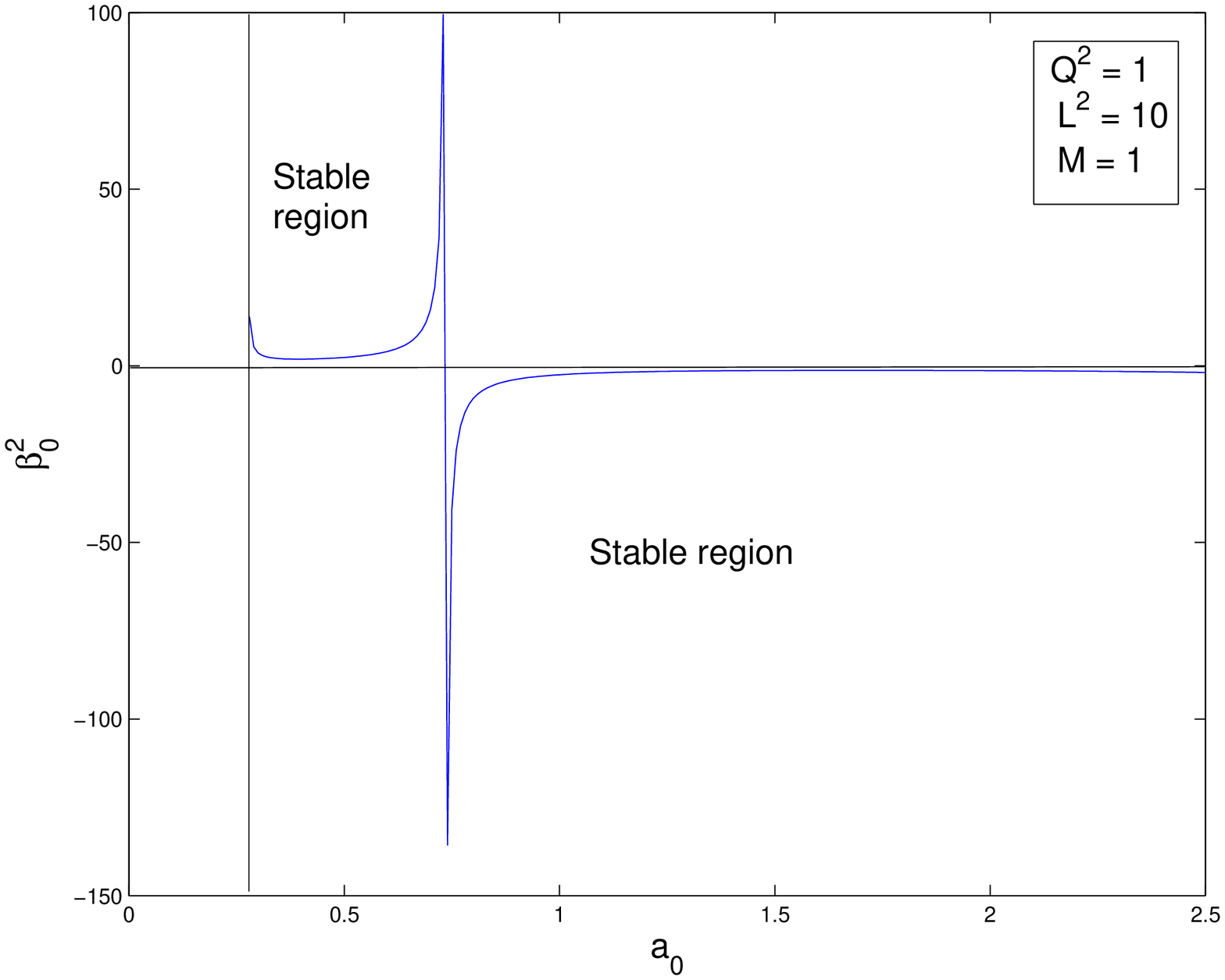}
\end{center}
\caption{ The region of stability is above the curve on the left and below the curve
on the right. } \label{fig5}
\end{figure}
\begin{figure}[ptb]
\begin{center}
\vspace{0.4cm}
\includegraphics[width=0.55\textwidth]{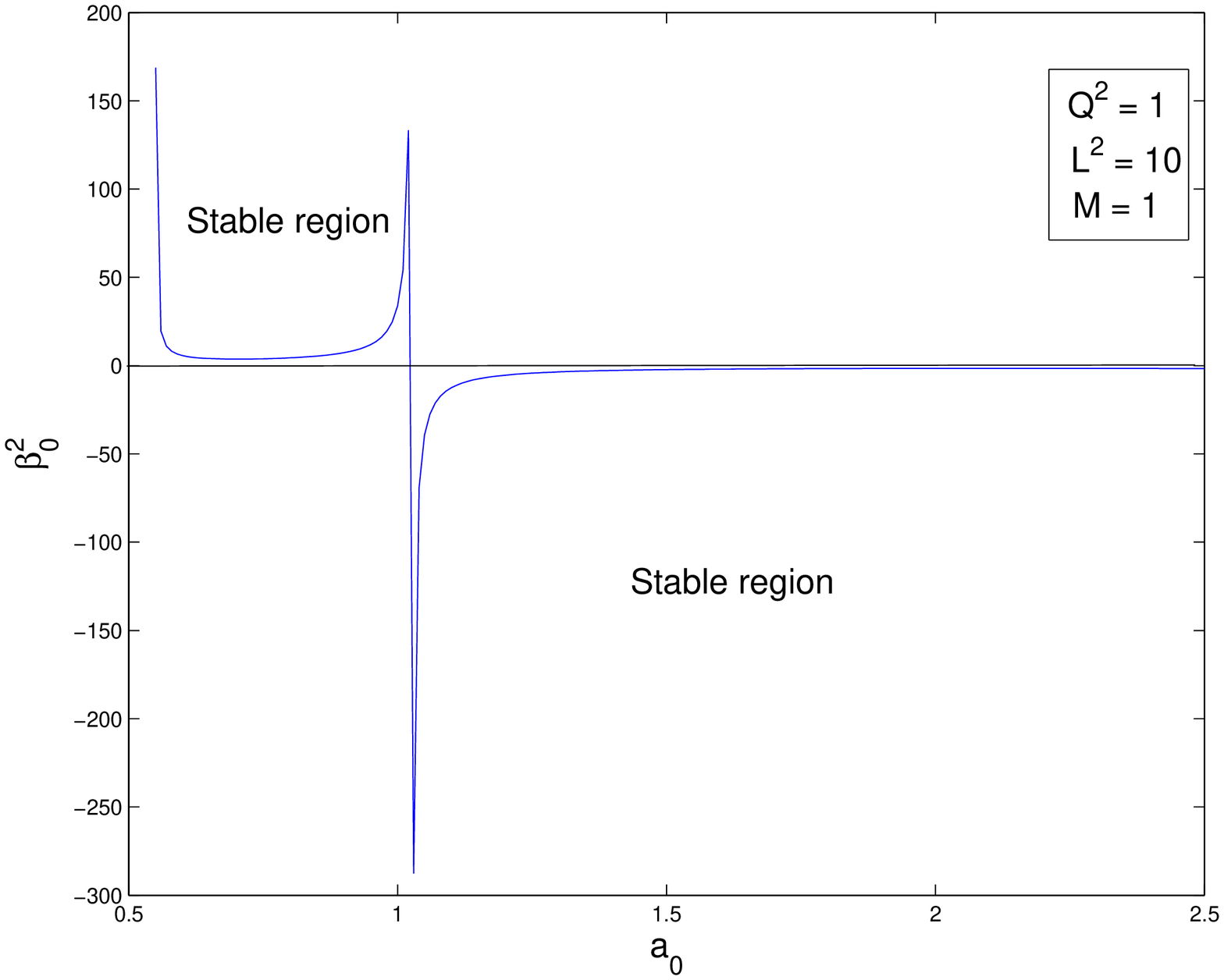}
\end{center}
\caption{ The region of stability is above the curve on the left and below the curve
on the right .} \label{fig5}
\end{figure}

\section{Conclusions}\noindent
In this work we construct charged thin-shell
wormhole from BTZ black hole using the cut-paste technique.
Though, a disadvantage with using BTZ in the thin-shell
wormholes construction is that it is not asymptotically
flat and therefore the wormholes are not asymptotically
flat. Therefore, there is some 'matter' at infinity.
There is also 'matter due to the charge'. Vacuum solutions,
or, at least asymptotically flat ones, are better because
of the 'good' behaviour at infinity.

   We obtain the surface stress energy tensor at the junction by
Lanczos equation where we observe that the energy density
$\sigma$ is negative but pressure $p$ is positive. Also we get $\sigma+p$ and $\sigma+2p$
positive which shows matter contained by the shell violates the
WEC but satisfy the NEC and SEC.

We draw our main attention on the stability of the shell
considering different equation of state for the exotic matter
at the throat. Firstly we consider dark, quintessence and phantom-like
equation of state changing the value of $w$, where we found
stable wormholes in all cases. Secondly we consider generalized Chaplygin gas
 equation of state and found stable wormhole for suitable choice
 of $\alpha$ with the help of graphical representation. Finally we studied  the
linearized stability analysis around the static solution i.e. at
$a=a_0$ and we found stable wormholes in suitable range
of $\beta^2_0$ with charged and mass though the behavior
of $\beta_0$ is not clear to us for exotic matter.

\subsection*{Acknowledgments}
AB is thankful to Dr. Farook Rahaman for this concept and helpful discussion.

\end{document}